\documentclass[twocolumn,prl,preprintnumbers,amsmath,amssymb]{revtex4}
\usepackage{amsmath,dcolumn,graphicx,epsf,ulem}% Include figure files
\setkeys{Gin}{width=8.6cm,keepaspectratio}
\usepackage{dcolumn}% Align table columns on decimal point \usepackage{bm}% bold math

\begin{document}

\title{Local Structure of $\rm\bf {La_{1-x}Sr_{x}CoO_3}$ determined
from EXAFS and neutron PDF studies}

\author{N.~Sundaram$^1$, Y.~Jiang$^1$, I.E.~Anderson$^1$, D.P.~Belanger$^1$,
C.H.~Booth$^2$, F.~Bridges$^1$, J.F.~Mitchell$^3$, Th.~Proffen$^4$,
H.~Zheng$^3$}

\affiliation{$^1$Department of Physics, University of California,
 Santa Cruz, CA 95064, USA}
\affiliation{$^2$Chemical Sciences Division, Lawrence Berkeley National Laboratory,
1 Cyclotron Rd.  Berkeley, CA 94720, USA}
\affiliation{$^3$Materials Science Division, Argonne National Laboratory,
9700 Cass Ave, Argonne, IL 60439, USA}
\affiliation{$^4$Lujan Neutron Scattering Center, Los Alamos National Laboratory,
Los Alamos, NM 87545, USA}

%\date{\today}
\begin{abstract}{ The combined local structure techniques, extended x-ray absorption fine
structure (EXAFS) and neutron pair
distribution function analysis, have been used for temperatures
$4 \le T \le 330$ K to 
rule out a large Jahn-Teller (JT) distortion of the
$\rm Co$-$\rm O$ bond in $\rm La_{1-x}Sr_{x}CoO_3$ for
a significant fraction of $\rm Co$ sites ($x \le 0.35$),
indicating few, if any, JT-active, singly occupied e$_g$ $\rm
Co$ sites exist.}

\end{abstract}


\maketitle

The cobaltite system $\rm La_{1-x}Sr_{x}CoO_3$ (LSCO) has a rich
temperature-concentration phase diagram as a result of multiple relevant energy
scales. In the parent compound, $\rm LaCoO_3$ (LCO), a significant thermal
population of an excited spin state occurs above $T=100$ K \cite{aynctshok94}.
Models generally include a low spin state (LS, $S=0$) at low
$T$, with electronic configuration ${t_{2g}}^6{e_g}^0$, along with
either an intermediate spin (IS, $S=1$), ${t_{2g}}^5{e_g}^1$, or a high spin
state (HS, $S=2$), ${t_{2g}}^4{e_g}^2$, that becomes populated as
$T$ increases.  Since the theoretical work of Potze {\it et al.}
\cite{psa95}, and Korotin {\it et al.} \cite{kesaks96}, which developed ideas
by Goodenough \cite{g71} and Zaanen {\it et al.} \cite{zsa85}, the IS state has
often been invoked for interpreting experimental results at intermediate
$T$ in LCO and LSCO.  More recently, the theoretical support for the
IS has been questioned \cite{kjhn06}. An excellent review is
given by Medarde, {\it et al.} \cite{mdgvppcntb06}. The IS state is expected to
be Jahn-Teller (JT) active, with local distortions of the $\rm O$ octahedra
surrounding the $\rm Co$ ions, breaking the 6-fold symmetry and creating $\rm
Co$-$\rm O$ bonds with different lengths. The related $\rm La_{1-y}Ca_yMnO_3$
(LCMO) manganite system exhibits a large JT distortion of the
$\rm Mn$-$\rm O$ bonds as reported in several recent extended x-ray absorption
fine structure (EXAFS) and neutron pair distribution function (PDF) studies
\cite{Downward2005,Jiang07,Bozin07}.  A similar distortion has been found in
$\rm La_{1-x}Sr_{x}MnO_3$ (LSMO) by Louca {\it et al.}, \cite{le99} (PDF),
while Mannella {\it et al.}, \cite{Mannella2004} (EXAFS) report a smaller
distortion for LSMO.  

If the proposed IS state is present in LCO, then it should
exhibit the associated JT distortion with a split $\rm Co$-$\rm O$ bond peak
for $T>100$ K but not for $T<<100$ K.  The
larger $\rm Sr$ ion decreases the crystal field splitting, which would be
expected to enhance the IS state \cite{spsp96}.  Previous magnetic
susceptibility work \cite{lstrk99} shows that in LSCO the excited
spin state persists to low $T$.  Hence, at low $T$ we should
see no split peak for LCO, followed by increasing evidence for
a splitting with increasing doping. More importantly, a
splitting should exist at $T=300$ K for all concentrations if the IS state is
the operative one. 

Previous neutron PDF studies of LSCO \cite{ls03}
indicated a local distortion comparable to LSMO with four short and
two long  $\rm Co$-$\rm O$ bonds. However, while some recent
experimental results \cite{ls03,gfycplsbb06}
are argued to
be consistent with a large JT distortion and its associated IS,
with the recent work of Klie {\it et al.} \cite{kzzvwl07} claiming to exclude
a transition to the HS state, others
\cite{mdgvppcntb06,psmmpcthk06,nkonm02} are argued to be
inconsistent with an LS-IS interpretation, but these are
not local structure measurements.  Here,
we report neutron PDF and EXAFS results which show that
the non-thermal local distortions in LSCO materials are,
in fact, much smaller than those in LCMO.

Powder samples of $\rm La_{1-x}Sr_{x}CoO_3$
($x=0$, $0.15$, $0.20$, $0.25$, $0.30$, and $0.35$) were
synthesized by Mitchell and Zheng at ANL
(MZ) and by Sundaram at UCSC (NS) using a standard
solid state reaction \cite{lstrk99}.
Stoichiometric amounts of $\rm La_2O_3$, $\rm
SrCO_3$ and $\rm Co_3O_4$ were ground thoroughly and fired several times at
temperatures ranging from $875$~$^\circ$C to $1100$~$^\circ$C for several days,
with a final heating at $1200$~$^\circ$C for one day in air.  X-ray diffraction
measurements confirmed the formation of pure
phase material and iodometric titrations or thermogravimetric analysis confirmed
correct oxygen stoichiometry.  Magnetization measurements
at the Lawrence Berkeley National Laboratory are consistent with previous
results \cite{ssttzps04}.

Temperature dependent EXAFS transmission $\rm Co$ $K$ edge data for the
LSCO samples were collected at the Stanford Synchrotron Radiation
Laboratory (SSRL). Fine powders were brushed onto tape; several tape layers were
used to make the step height (the x-ray absorption increase at the $\rm Co$
$K$ edge) 0.3 $\sim$ 0.5. The double $\rm Si$ (111) monochromator was detuned 50\%
to reduce harmonics.  The energy resolution was $\sim$$1$~eV.

A standard data reduction was used to extract the EXAFS $k$-space
oscillations, which were Fourier transformed into $r$-space (RSXAP package
\cite{RSXAP}). Next, the data were fit to theoretical EXAFS functions generated
by FEFF 8.20 (Rehr and co-workers \cite{FEFF8}), using the program \verb|rsfit|
(RSXAP package). Our primary interest here is $\sigma$, the width of the $\rm
Co$-$\rm O$ PDF, which parametrizes the amount of distortion present around
the $\rm Co$ ions.  Note that uncorrelated contributions to $\sigma^2$ add in
quadrature; {\it i.e.} $\sigma_{total}^2$
= $\sigma_{static}^2$ + $\sigma_{phonons}^2$ + $\sigma_{Jahn-Teller}^2$.

The thermal phonon contributions were determined from a fit of $\sigma^2(T)$
vs.\ $T$, for $4 \le T \le 330$~K, to the correlated Debye model plus a static
off-set. This model is usually a good approximation for all phonon modes
\cite{Ashcroft76} including acoustic and optical phonons
\cite{Lee77,Teo86}; see Ref.\ \cite{Jiang07} for details about our
use of this model.

\begin{figure}

\includegraphics[width=2.7in,angle=0]{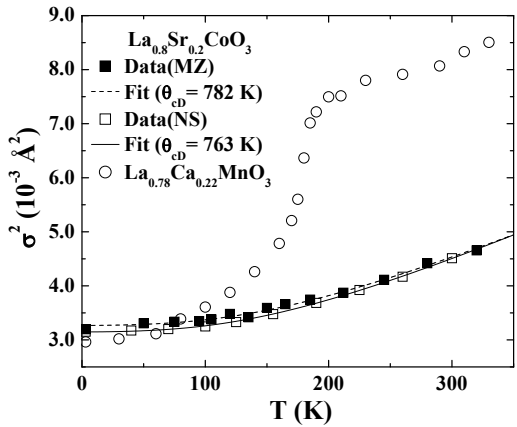}

\caption{$\sigma^2(T)$ for the $\rm La_{0.8}Sr_{0.2}CoO_3$ MZ
and NS samples, and correlated Debye fits.
The data overlap very well within
the relative errors that are comparable or smaller than the symbol sizes.
Similar results for a $\rm
La_{0.78}Ca_{0.22}MnO_3$ sample ($\rm Mn$-$\rm O$ bond,
see Ref.\ \onlinecite{Jiang07})
are replotted here; this sample has a large JT distortion
that develops between $100$ and $200$~K.}

\label{debye}
\end{figure}

\begin{figure}
\includegraphics[width=2.7in, angle=0]{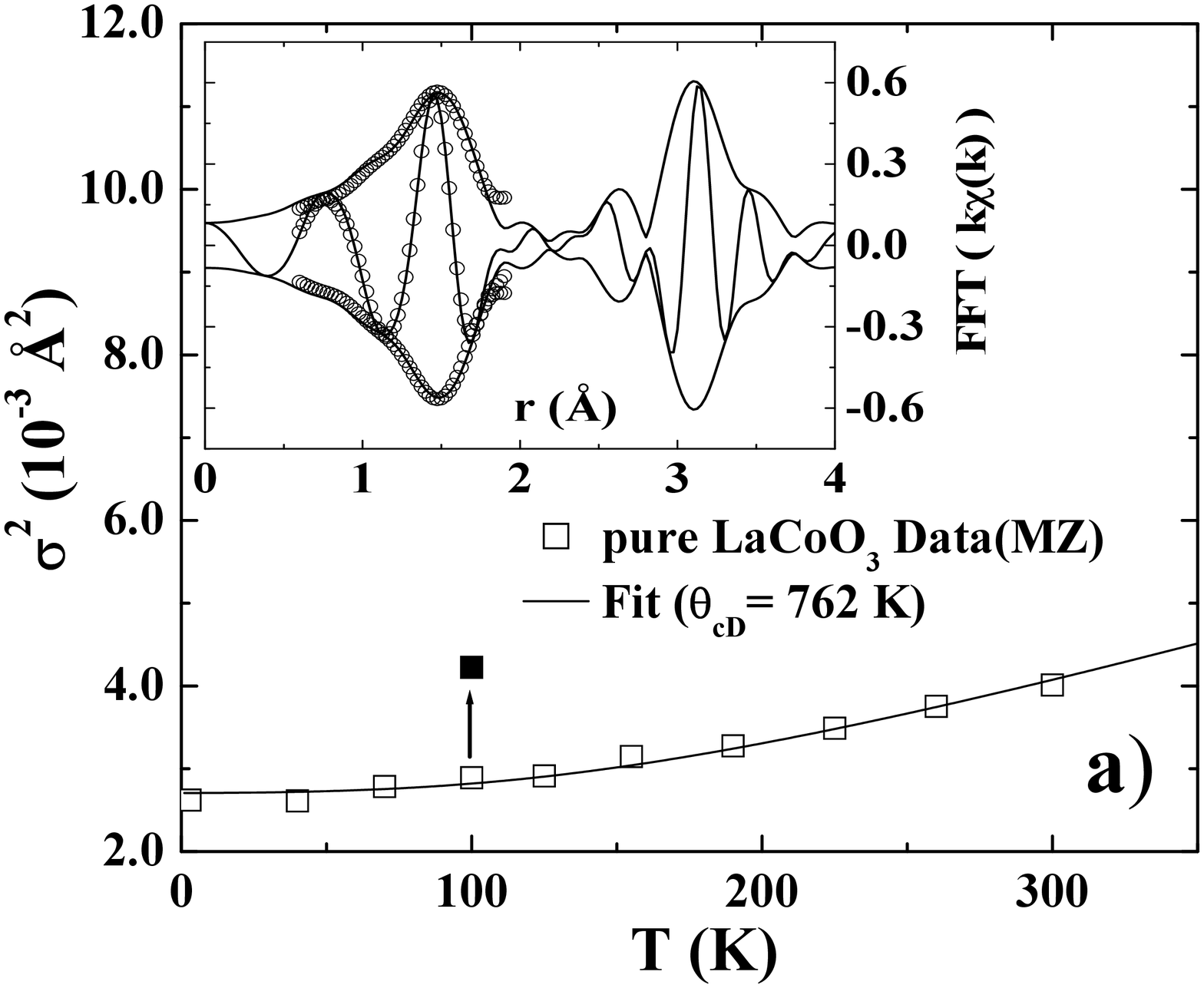}
\includegraphics[width=2.7in, angle=0]{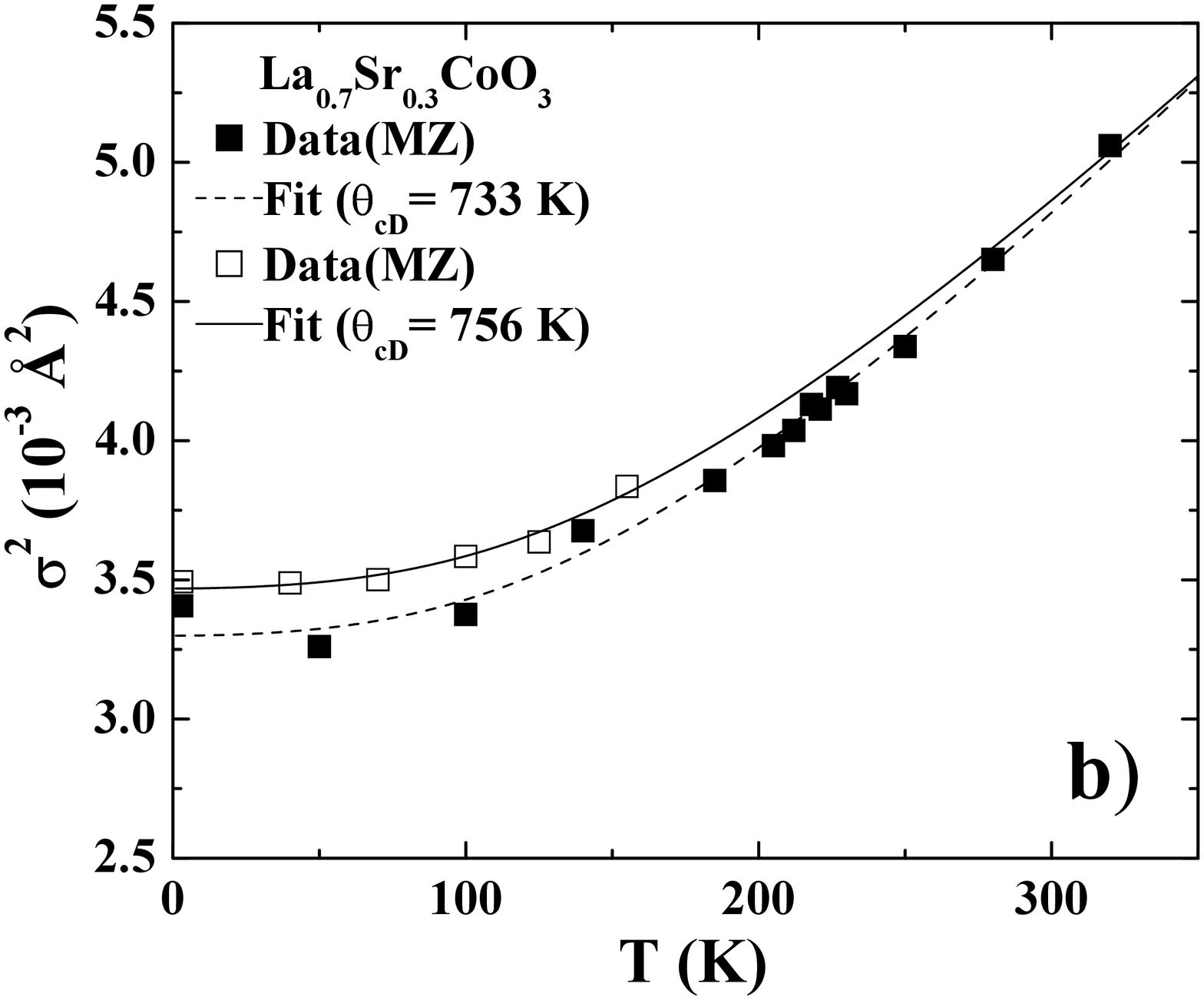}

\caption{a) $\sigma^2(T)$ for $\rm LaCoO_3$ and the correlated Debye fit.
The inset is an example of $4$~K $r$-space data (solid line) and a fit to
$\rm Co$-$\rm O$ first peak(open circle); the Fourier Transform (FT) range is
$3.3-12.0$ \AA$^{-1}$ with a Gaussian broadening of $0.3$ \AA$^{-1}$, and
the r-space fit range is $1.1$ to $1.7$ \AA \ .  Note that we are fitting to the
fast oscillating real and imaginary (not shown) parts of the FT. A good fit is
obtained from $0.7 - 1.8$ \AA. b) $\sigma^2(T)$ for the $\rm
La_{0.7}Sr_{0.3}CoO_3$ JM samples. The solid squares represent the data from
the first measurement, and open squares represent the data from the second
measurement showing consistency between measurements taken a year apart; the
dashed line is the correlated Debye fit for the first experiment data, while
the solid line is the fit for the second experiment data. The solid symbol in
a) shows the increase in $\sigma^2$ if a 3-peak splitting occurs near
$100$~K.}

\label{bulksig2}
\end{figure}

In Fig.\ \ref{debye} we plot $\sigma^2(T)$ and the correlated Debye fits for MZ
and NS $\rm La_{0.8}Sr_{0.2}CoO_3$ samples. The relative errors (Figs. 
\ref{debye} and \ref{bulksig2}1) are comparable to the scatter in the data;
systematic errors, which should be the same at all $T$ for a given
sample and nearly the same for all samples, are also present. Such errors,
which shift the entire plot up or down on the vertical axis, are estimated to
be less than $5\times 10^{-4}$ \AA$^2$.
We also plot, for comparison,
corresponding results for a 22\% $\rm Ca$-doped LCMO sample \cite{Jiang07}
which has a metal-insulator (MI) transition around $190$~K associated with a
local JT distortion of the $\rm Mn$-$\rm O$ octahedra. Between $100$
and $200$~K, the JT splitting results in a configuration with two longer $\rm
Mn$-$\rm O$ bonds and four relatively shorter bonds for some sites (at low $T$,
the $\rm Mn$-$\rm O$ octahedron has six bonds of nearly equal length). The
proposed JT distortion of the $\rm Co$-$\rm O$ bonds in
LSCO \cite{ls03} is
of comparable magnitude to that observed in LCMO. However, the lack of any
significant step for LSCO in Fig.\ \ref{debye} and the rather small static
distortion at 4K ($\le$ 0.0006 \AA$^2$) indicates very little JT
distortion of the $\rm Co$-$\rm O$ PDF peak between $4$ and $300$~K for the
20\% $\rm Sr$ doped LSCO samples. One cannot, of course, rule out the
possibility of a few percent of sites having a JT distortion.

Fig. \ref{bulksig2} shows $\sigma^2(T)$ ($\rm Co$-$\rm O$ peak) for two
samples. 
The plots indicate that
LCO has a slightly less distorted environment around the $\rm Co$
atoms; however, the difference in $\sigma^2$ at $4$~K between this sample and
the others is comparable to the systematic uncertainty, so it is not a
definitive result. For the 30\% Sr sample [Fig.\ \ref{bulksig2}(b)], data
from one experiment overlap well with those from an experiment one year
earlier. These results together with the 20\% LSCO data in Fig.\ \ref{debye}
show the consistency of the data collection and analysis.

The correlated Debye temperature ($\theta_{cD}$) for all samples is
approximately $760$~K and is a measure of the Co-O bond strength. For this relatively high value of $\theta_{cD}$ and the
limited $T$ range (T/$\theta_{cD}$ $<$ 0.4), $\theta_{cD}$ is strongly
dependent on small fluctuations of the $\sigma^2(T)$ data points, particularly
at higher $T$; thus this relatively large observed scatter in
$\theta_{cD}$ (see Figs.\ \ref{debye},\ref{bulksig2}) is expected and the
overall uncertainty is likely $30$-$40$ K. For comparison $\theta_{cD}$ for 
the Mn-O peak in LCMO is $\sim$ 830 K \cite{Jiang07}. 
The static contribution,
$\sigma^2_{static}$, obtained from these Debye fits, range from
$0.3$ - $7.8\times
10^{-4}$ \AA$^2$ for all our samples ($0 \le x \le 0.3$).  The
zero-point-motion contribution to $\sigma^2$ for this value of $\theta_{cD}$ is
2.6$\times$10$^{-3}$ \AA$^2$. Hence, this static contribution is very small for
all samples and consistent with $\sigma^2_{static}$ = 0 within our 
systematic uncertainty, $5\times 10^{-4}$ \AA$^2$. 

There are three published EXAFS papers on the $\rm Sr$ doped cobaltites;
however, they only report room temperature data
\cite{Sikolenko06,Haas04}. When fitting the first shell $\rm Co$-$\rm
O$ peak to a single Gaussian, they all obtain similar $\sigma^2$(300 K) results to those we report
above. However, a diffraction study has suggested that a 3-peak splitting of
the $\rm Co$-$\rm O$ bonds develops above $\sim$ 70 K (space group change from
$R\overline{3}c$ to $I2/a$, $\Delta r$ $\sim$ 0.05-0.06 \AA, although the change
in the fit parameter for the two fits is very small)\cite{Maris03}.  Two of the
EXAFS groups therefore tried a split-peak fit and suggest a splitting
of approximately $0.06$ \AA \ \cite{Haas04}, much smaller than
the $0.15$ \AA{ } splitting reported by Louca {\it et al.}, \cite{le99}.
However, $0.06$ \AA \ is well below the Fourier limit for resolving a split
peak in EXAFS when the maximum $k$ is $\sim$ 13-14 \AA$^{-1}$; a split peak
cannot be resolved if $\Delta$r $<$ $\frac{\pi}{2k_{max}}$\ \cite{Lee81,Teo86}.

For comparison with these other reports, we have also included the effect of a
splitting of the $\rm Co$-$\rm O$ peak into 3-peaks between $60$ and $100$ K.
In that case for peaks at $r$, $r$$\pm \Delta r$, there is an additional static
contribution - $\sigma^2_{static}$ = 2/3{($\Delta r)^2$ \cite{Teo86}.  Using
$\Delta r$ = 0.05 \AA{ } and the zero-point-motion contribution above, the value
of $\sigma^2$ would increase to 4.2$\times$10$^{-3}$ \AA $^2$, as
shown by a solid point at 100K in Fig.\ \ref{bulksig2}(a); a simulation made using three such
peaks, and fit using a single peak as for the data, yielded the same result.
We would easily see a step change of this magnitude - it is inconsistent with
the EXAFS data.  An upper limit to a 3-peak splitting, using a possible step
change in $\sigma^2$ that is three times the scatter in the data, is $\Delta r
\leq$ 0.025 \AA.  Also if we assume a 3-peak JT splitting occurs at all T, an
upper limit for such a splitting from the static contribution to $\sigma^2$ is
$\Delta r \leq$ 0.03 \AA; assuming a 2-peak splitting model instead, the maximum
$\Delta r$ would be  $\approx 0.05$ \AA.  Hence, any JT distortion at low $T$
must be small, or the fraction of JT split sites less than $5$-$10$\%.  

\begin{figure}
\includegraphics[angle=270,width=2.6in,angle=0]{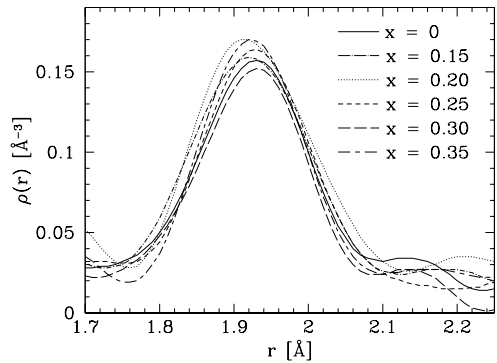}

\caption{$\rho (r)$ for all the samples at $T=300$ K with
$Q_{max}=35$ \AA$^{-1}$.  The large peak represents a $\rm Co$-$\rm O$
bond length of $1.92$ \AA. There is little evidence for the peak
previously reported \cite{le99} at $2.15$ \AA.}

\label{sim}
\end{figure}

The neutron PDF technique is another local structure probe well suited for
investigating JT splittings of the $\rm Co$-$\rm O$ bonds \cite{eb03}, as long
as the splitting is larger than the resolution and involves a significant
number of Co sites. As discussed above, if the proposed IS state exists, then
the LCO should exhibit a JT distortion with a split
$\rm Co$-$\rm O$ bond peak for $T>100$ K but not for
$T<<100$ K, while the doped compounds should show a splitting at
all $T$.  We collected neutron diffraction data for
the NS samples using the high-resolution neutron powder diffractometer NPDF
\cite{pebclp02} at the Lujan Neutron Scattering Center, LANL.
Data at $T=12$, $100$, and $300$ K for each sample
were corrected for detector dead time and efficiency, background,
absorption, multiple scattering, and inelastic effects. The data were then
normalized by the incident flux and the total sample scattering cross section
to yield the total scattering structure function $S(Q)$, from which the PDF,
$G(r)$, and the pair density function, $\rho(r)$, were obtained by a Fourier
transform \cite{eb03}. Data were collected with reciprocal lattice
vectors as large as
$Q=45$ \AA$^{-1}$, giving a high real-space resolution $\le$
$0.1$ \AA.  The program PDFGETN \cite{pgpb00} was used for data processing.

Figure \ref{sim} shows no evidence for a split peak in the PDF
data from any sample at $T=300$ K, nor is one observed at lower
$T$. The small peak that seems to appear near $2.15$ $\sim$ $2.2$ \AA
\ is much smaller than the peak observed in earlier PDF measurements
\cite{le99} at $2.1$ \AA, and it shifts in $r$ with changing values of
$Q_{max}$, the maximum cutoff in $Q$, as exemplified by three different
$Q_{max}$ in Fig.\ \ref{simQ} for the same set of data. The strong $Q_{max}$
dependence demonstrates that the small peak does not represent a splitting
from the main peak and hence, a JT distortion, but is primarily an effect of
termination ripples \cite{eb03} introduced in the Fourier transform by changes
in $Q_{max}$.
Peaks representing real structure are not
sensitive to $Q_{max}$. It is clear that the position of
the large $\rm Co$-$\rm O$ peak at $1.92$ \AA \ does not significantly change
position.

\begin{figure}
\includegraphics[angle=270,width=2.6in,angle=0]{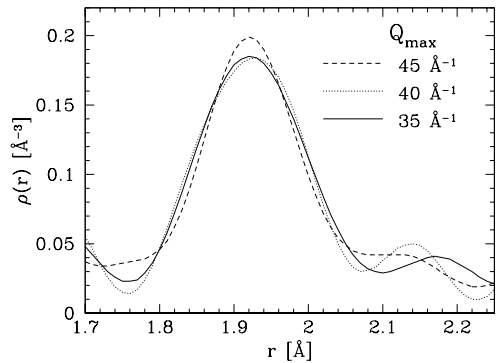}

\caption{$\rho(r)$ obtained from the scattering data on LCO using different
$Q_{max}$ at $T=200$ K.  Although a small peak can be seen between $2.1$ and
$2.2$ \AA, its position changes with $Q_{max}$
and does not indicate a real bond length in this $r$ range.}

\label{simQ}
\end{figure}

Recent neutron PDF studies \cite{plkhy07} of
$\rm {La_{1-y}Ba_{y}CoO_3}$ and $\rm {La_{1-y}Ca_{y}CoO_3}$ 
indicated asymmetric $\rm Co$-$\rm O$ peaks, particularly in the
former compound, giving evidence for the IS state in these compounds.
We observed no evidence of asymmetry in our LSCO data.

Rietveld refinements using the $I2/a$ space group reported by Maris {\it et al.}
\cite{Maris03} did not yield better fits of our data than the $R\overline{3}c$
space group used by most other groups.  Moreover, the three $\rm Co$-$\rm O$
bond lengths found are well within $0.001$ \AA{} of each other.  Using the
Maris {\it et al.} results with a split first $\rm Co$-$\rm O$ peak, we calculated
the corresponding PDF; the results are not consistent with our data. We
conclude that our data are fit better by the $R\overline{3}c$ space group,
as will be reported elsewhere.

Our neutron data show no evidence for a JT distortion in LCO or LSCO, in
complete agreement with the EXAFS results described earlier.  Together, our
EXAFS and neutron PDF data indicate that, if the localized spin model with the
LS $\rightarrow$ IS scenario is operative, either the $e_g$ state does not
couple to the lattice as much as previously expected, resulting in undetectably
small distortions, or it involves very few $\rm Co$ sites. This result should be
considered in the interpretation of the low energy excitations observed in LSCO
and LCO \cite {plrlqcozmcsm06}.  Recent experiments have been interpreted as
being more compatible with a LS $\rightarrow$ HS scenario, perhaps with
inhomogeneous mixed-spin states and strong $\rm Co$-$\rm O$ orbital
hybridization \cite{psmmpcthk06,mdgvppcntb06,c99}. Our combined
local structure results
remove an apparent contradiction between such measurements and earlier 
local structure results\cite{ls03}.   An
important consideration is the hole hopping rate in conducting samples - if the
hole hops faster than phonon time scales, the $\rm O$ atoms have no time to
respond and no (or little) distortion will be observed: for example, very
little distortion is observed for the CMR manganites at low $T$ while
a small JT distortion is observed in LSMO at high $T$ where the
polarons are hopping rapidly\cite{Mannella2004}.  Using the
localized spin models may be incomplete if the conducting charges have a
wavefunction spread over several sites \cite{mklamipma07}.

More extensive studies of $\rm {La_{1-x}Sr_{x}CoO_3}$ will be reported
elsewhere, including results for bulk powders, such as those used here, and
$20$-$40$~nm nanoparticle powders.

\acknowledgments

Work at UCSC was partly funded by DOE Grant No.\
DE-FG02-05ER46181.  Work at Argonne National Laboratory was supported by
the US DOE, Office of Science, under Contract No.
DE-AC-02-06CH11357. EXAFS experiments were performed at SSRL (operated by the
DOE, Div. of Chemical Sciences, and by the NIH, Biomedical Resource
Technology Program, Div. of Research Resources).
This work has benefited from the use of NPDF at the Lujan Center, funded by DOE Office of Basic Energy Sciences (BES).
Los Alamos National Laboratory is operated by Los Alamos National
Security LLC under DOE Contract DE-AC52-06NA25396.  Work at Lawrence Berkeley
National Laboratory was supported by U. S. DOE, BES,
under contract DE-AC02-05CH11231.


\begin{thebibliography}{999}

\bibitem{aynctshok94}K.~Asai {\it et al.},
Phys.\ Rev.\ B {\bf 50}, 3025 (1994).

\bibitem{psa95}R. H. Potze, G.A.~Sawatzky, and M.~Abbate,
Phys.\ Rev.\ B {\bf 51}, 11501 (1995).

\bibitem{kesaks96}M.A.~Korotin {\it et al.},
Phys.\ Rev.\ B {\bf 54}, 5309 (1996).

\bibitem{g71}J.B. Goodenough,
Mat. Res. Bull. {\bf 6}, 967 (1971).

\bibitem{zsa85}J. Zaanen, G.A.~Sawatzky, and J.W.~Allen,
Phys. Rev. Lett. {\bf 55}, 418 (1985).

\bibitem{kjhn06}K. Knizek {\it et al.},
J. Phys.: Condens.\ Matter {\bf 18}, 3285 (2006);
R. Radwanski and Z. Ropka,
Sol.\ State Commun. {\bf 112}, 621 (1999);
R. Radwanski and Z. Ropka,
Physica B {\bf 281-282}, 507 (2000);
Z. Ropka and R.J. Radwanski,
Phys.\ Rev.\ B {\bf 67}, 172401 (2003).

\bibitem{mdgvppcntb06}M. Medarde {\it et al.},
Phys.\ Rev.\ B {\bf 73}, 54424 (2006).

\bibitem{Downward2005}L.~Downward {\it et al.},
Phys.\ Rev.\ Lett.\ {\bf 95}, 106401 (2005).

\bibitem{Jiang07} Y.~Jiang {\it et al.},
Phys.\ Rev.\ B {\bf 76}, 224428 (2007).

\bibitem{Bozin07}E.S.~Bozin {\it et al.},
Phys.\ Rev.\ Lett.\ {\bf 98}, 137203 (2007).

\bibitem{le99}D. Louca and T. Egami,
Phys.\ Rev.\ B {\bf 59}, 6193 (1999).

\bibitem{Mannella2004}N.~Mannella {\it et al.},
Phys.\ Rev.\ Lett.\ {\bf 92}, 166401 (2004).

\bibitem{spsp96}V.G. Sathe {\it et al.},
J. Phys.: Condens. Matter {\bf 8}, 3889 (1996).

\bibitem{lstrk99}D.~Louca {\it et al.},
Phys.\ Rev.\ B {\bf 60}, 10378 (1999).

\bibitem{ls03}D.~Louca and J.L.~Sarrao,
Phys.\ Rev.\ Lett.\ {\bf 91}, 155501 (2003);
D. Phelan {\it et al.},
Phys.\ Rev.\ Lett. {\bf 97}, 235501 (2006).

\bibitem{gfycplsbb06}V. Gnezdilov {\it et al.},
Low Temp.\ Phys. {\bf 32}, 162 (2006);
G. Vanko {\it et al.},
Phys.\ Rev.\ B {\bf 73}, 24424 (2006);
V.P. Plakhty {\it et al.},
J. Phys.: Condens.\ Matter {\bf 18} 3517 (2006);

\bibitem{kzzvwl07}R.F. Klie {\it et al.},
Phys.\ Rev.\ Lett. {\bf 99}, 47203 (2007).

\bibitem{psmmpcthk06}A.~Podlesnyak {\it et al.},
Phys.\ Rev.\ Lett. {\bf 97}, 247208 (2006);
M.W. Haverkort {\it et al.},
Phys.\ Rev.\ Lett. {\bf 97}, 176405 (2006).

\bibitem{nkonm02}S. Noguchi {\it et al.},
Phys.\ Rev.\ B {\bf 66}, 94404 (2002).

\bibitem{ssttzps04}V.V. Sikolenko {\it et al.},
J. Phys.: Condens. Matter {\bf 16}, 7313 (2004).

\bibitem{RSXAP}C.H. Booth R-Space X-ray Absorption Package,
http://lise.lbl.gov/RSXAP.

\bibitem{FEFF8} A.L.~Ankudinov {\it et al.},
Phys.\ Rev.\ B {\bf 58}, 7565 (1998).

\bibitem{Ashcroft76} N.W.~Ashcroft and N.D.~Mermin, in Solid State
Physics (Saunders College, Philadelphia, 1976).

\bibitem{Teo86} B.K.~Teo, EXAFS: Basic Principles and
Data analysis (Springer-Verlag, Ney York, 1986).

\bibitem{Lee77} P.A.~Lee and G.~Beni,
Phys.\ Rev.\ B {\bf 15}, 2862 (1977);
A.~Bianconi, in {\it X-ray Absorption: principles,
applications, techniques of EXAFS, SEXAFS and XANES}, edited
by D.~C.~Koningsberger and R.~Prins (John Wiley and Sons,
New York, 1988), p. 594.

\bibitem{Sikolenko06} V.V. Sikolenko {\it et al.},
Cryst. Rep. {\bf 51}, S67 (2006).

\bibitem{Haas04} O.~Haas {\it et al.},
J. Solid State Chem. 177, 1000 (2004);
S.K. Pandey {\it et al.},
J. Phys.: Condens. Matter {\bf 18}, 10617 (2006).

\bibitem{Maris03}G.~Maris {\it et al.},
Phys.\ Rev.\ B {\bf 67}, 224423 (2003).

\bibitem{Lee81}P.A.~Lee, P.H.~Citrin,
P.~Eisenberger and B.M.~Kincaid,
Rev.\ Mod.\ Phys.\ {\bf 53}, 769(1981).

\bibitem{eb03}T.~Egami and S.J.L.~Billinge,
{\it Underneath the Bragg Peaks: Structural Analysis of Complex Materials},
Pergamon Materials Series, Volume 7, (Pergamon, New York, 2003i).

\bibitem{pebclp02}Th.~Proffen {\it et al.},
Appl.\ Phys.\ {\bf A74}, S163 (2002).

\bibitem{pgpb00}P.F.~Peterson {\it et al.},
J. Appl.\ Cryst.\ {\bf 33}, 1192 (2000).

\bibitem{plkhy07}D.~Phelan {\it et al.},
Phys.\ Rev.\ B {\bf 76}, 104111 (2007).

\bibitem{plrlqcozmcsm06}D. Phelan {\it et al.},
Phys.\ Rev.\ Lett. {\bf 96}, 27201 (2006);

\bibitem{c99}R.~Caciuffo {\it et al.},
Phys.\ Rev.\ B {\bf 59}, 1068 (1999).

\bibitem{mklamipma07}I.I. Mazin {\it et al.},
Phys.\ Rev.\ Lett. {\bf 98}, 176406 (2007).

\end{thebibliography}
\end{document}